\documentstyle[12pt]{article}

\newtheorem{thm-def}{Definition-Theorem}
\newtheorem{thm}{Theorem}
\newcommand{\s}{\sigma}
\renewcommand{\t}{\tau}
\newcommand{\p}{product integral}

\newcommand{\beq}{\begin{equation}}
\newcommand{\eeq}{\end{equation}}

\title{\nopagebreak
\begin{flushright}
\tenrm UCTP117.99
\end{flushright}\vskip0.3in
\nopagebreak
\large \bf  Product Integral Formalism \\
and \\  Non-Abelian Stokes Theorem}
\author{Robert L. Karp\thanks{e-mail address:
karp@physics.uc.edu},\,
Freydoon Mansouri\thanks{e-mail address: mansouri@uc.edu}\\
{\it \small
Physics Department, University of
Cincinnati, Cincinnati, OH 45221}\\
Jung S. Rno\thanks{e-mail address: rno@uc.edu} \\
{\it \small
Physics Department, University of Cincinnati-RWC, Cincinnati, OH
45236}}
\date{}

\begin{document}
\maketitle

\begin{abstract}
We make use of the properties of product integrals to obtain a
surface product integral representation for the
Wilson loop operator. The result can be interpreted as the
non-abelian
version of Stokes theorem.

{\bf PACS}: 11.15, 12.10, 12.15, 12.38.A
\end{abstract}

\section{Introduction}
The
main purpose of this work is to make use of
product integrals
to give two unambiguous proofs of the non-abelian
version of
Stokes theorem.
The product integral formalism has been used extensively in the
theory of
differential equations and of matrix valued
functions~\cite{rfour}. In the
latter context, it has a built-in feature for keeping track of
the {\it
order} of the matrix valued functions involved. As a result,
product
integrals are ideally suited for the description of {\it path
ordered}
quantities such as holonomies. 
Moreover, since the theory
is well developed independently of particular
applications, we
can be confident that the properties of such path ordered
quantites
which we establish using this method are correct and unambiguous.

Among the important advantages of the product integral
representation of the path dependent exponential of a matrix
valued function is that in such a framework
the Banach space structure of the corresponding matrix valued
functions is already built into the formalism. In particular, for
a closed path eclosing an orientable 2-surface, this will permit
a surface representation of such operators. 
Based on the central role of
Stokes
theorem in physics and in mathematics, it is not surprising that
the
non-abelian version of this theorem has attracted a good deal of
attention
in the physics literature~\cite{rfive}-~\cite{r15}. The central
features
of the earlier attempts\cite{rfive}-~\cite{r11} have been
reviewed and
improved upon in a recent work~\cite{r12}. Other recent works on
non-abelian Stokes theorem~\cite{r13,r14,r15} focus on specific
problems
such as confinement~\cite{r14}, zig-zag symmetry~\cite{r15}
suggested by
Polyakov~\cite{r16}, etc. With one exception~\cite{r12}, the
authors of
these works seem to have been unaware of a 1927 work in the
mathematical
literature by Schlessinger~\cite{r17}, which bears strongly on the
content
of this theorem. Schlesinger's work dealt with integrals of
matrix
valued
functions and their ordering problems. This amounts to
establishing
the non-abelian Stokes
theorem
in two (target space) dimensions. By an appropriate extension and
reinterpretation of his results, we show using the product
integral
approach that this theorem is valid in any target
space
dimension.

This work is organized as follows: To make this manuscript
self-contained,
we review in Section II the main features of product
integration~\cite{rfour} and state without proof a number of
theorems
which will be used in the proof of the non-abelian Stokes
theorem. In Section III, we deal with path ordered
exponentials of matrix valued fuctions which can be expressed as
product integrals and turn to the proof of the non-abelian
Stokes theorem for orientable surfaces. In section IV, we give a
variant
of this proof. In
section V, we explicitly demonstrate the gauge covariance of the
results obtained in sections III and IV.
Finally,
section is devoted to some additional remarks.

\section{Some properties of product integrals}
One of the initial motivations for the introduction of product
integrals was~\cite{rfour} to solve differential equations of the
type
\beq
Y'(s) =A(s)Y(s).
\eeq
In this expression, $Y(s)$ is an n-dimensional vector, $A(s)$ is
a matrix valued function, and prime indicates differentiation.
So, for two real numbers $a$ and $b$, the problem is to obtain
$Y(b)$  given $Y(a)$.
To deal with this problem, we make a partition
$P=\{s_0,s_1,\ldots,s_n\}$ of
the interval $[a,b]$. Let $\Delta s_k=s_k-s_{k-1}$ for
$k=1,\ldots,n$, and set $a = s_0$, $b = s_n$. Then, solving the
differential equation in each subinterval, we can write
approximately~\cite{rfour}
\beq
Y(b)\approx
\prod_{k=1}^{n}e^{A(s_k)\Delta s_k}Y(a)
\equiv \Pi_p (A) Y(a).
\eeq
Since $A(s)$ is matrix valued, the order in this product is
important.
Let $\mu(P)$ be the length of the longest $\Delta s_k$ in the
partition $P$. Then, as $\mu (P) \rightarrow 0$, we get
\beq
Y(b)=\lim_{\mu(P)\rightarrow 0} \Pi_P(A) Y(a).
\eeq
The limit is clearly valid for all $Y(a)$.

The limit of the ordered product on the right hand side of Eq.(3)
is the fundamental expression in the definition of a product
integral~\cite{rfour}. It is formally defined, in an obvious
notation, as
\beq
\prod_{a}^{x} e^{A(s)ds}
=\lim_{\mu(P)\rightarrow 0} \Pi_P(A) \equiv F(x,a).
\eeq
It is easy to see that $F(x,a)$ satisfies the differential
equation
\beq
\frac{d}{dx}F(x,a)=A(x)F(x,a)
\eeq 
with $ F(a,a)=1$.
The corresponding integral equation is
\beq
F(x,a)=1+\int_{a}^{x}\,ds\,A(s)F(s,a).
\eeq
Clearly, $F(a,a) = 1$, and $F(x,a)$ is unique.

Consider now some of the properties of the product integral
matrices. For each $x \epsilon [a,b]$ the product integral is
non-singular, and its determinant is given by
\beq
\det\left(\prod_{a}^{x} e^{ A(s)ds}\right)=e^{ \int_{a}^{x}\,
trA(s)ds}.
\eeq
In analogy with the additive property of ordinary integrals,
product integrals have the multiplicative property, or the
composition rule, 
\beq
\prod_{z}^{x} e^{ A(s)ds}=\prod_{y}^{x} e^{ A(s)ds}\prod_{z}^{y}
e^{A(s)ds}.
\eeq
Where $x, y, z \epsilon [a,b]$ and $z \leq y \leq x$. The result
is independent of the choice of $y$ and any further decomposition
of the products on the right hand side. 

Derivatives
with respect to the end points are given by
\beq \frac{\partial}{\partial x}\prod_{y}^{x} e^{ A(s)ds}
=A(x)\prod_{y}^{x} e^{ A(s)ds}, \eeq
and
\beq
\frac{\partial}{\partial y}\prod_{y}^{x} e^{ A(s)ds} =
-\prod_{y}^{x} e^{ A(s)ds} A(y).
 \eeq
One of the fundamental features associated with a connection is
the notion
of parallel transport. To see how it can be formulated in product
integral formalism, consider a map $P : [a,b] \rightarrow {\bf
C}_{n
\times n}$, which is continuously differentiable. Then $P(x)$ is
an indefinite product integral if for a given $A(s)$
\beq
P(x) = \prod_{a}^{x} e^{ A(s)ds} P(a).
\eeq

Next, we define an operation known as $L$ operation which is
like the logarithmic derivative operation on non-singular
functions. Let
\beq LP(x) = P'(x)P^{-1}(x),
\eeq
where prime indicates differentiation. Then, from Eq. (11) it
follows that
$$(LP)(x) = A(x)$$.
One of the byproducts of this operation is that
\beq
L(PQ)(x)=LP(x)+P(x)(LQ(x))P^{-1}(x).
\eeq
The $L$ operation is a crucial ingredient in establishing the
analog of the fundamental theorem of calculus for product
integrals. With the map $P$ as defined above, this theorem states
that
\beq
\prod_{a}^{x} e^{ (LP)(s)ds}=P(x)P^{-1}(a).
\eeq
From the results given above, it follows that $P$ is a solution
of the initial value problem
\beq
P'(x) = (LP)(x) P(x).
\eeq
With the unique solution given by Eq. (11), this establishes the
fundamental theorem of product integration. Just as in ordinary
integration, the knowledge of simple product integrals can be
used
to evaluate more complicated product integrals. For example, one
can prove the {\it sum rule} for product integrals:
\beq
\prod_{a}^{x} e^{ [A(s)+B(s)]ds}=P(x)\prod_{a}^{x} e^{
P^{-1}(s)B(s)P(s)ds}.
\eeq
Finally, we state two other important properties of product
integrals which will be used in the sequel. One is the {\it
similarity
theorem} which states that
\beq
P(x)\prod_{a}^{x} e^{ B(s)ds}P^{-1}(a)=\prod_{a}^{x} e^{
[LP(s)+P(s)B(s)P^{-1}(s)]ds}.
\eeq
The other property is differentiation with respect to a
parameter. Let
\beq
P(x,y;\lambda)=\prod_{y}^{x} e^{ A(s;\lambda)ds},
\eeq
where $\lambda$ is a parameter. Then the differentiation rule with
respect to this parameter is given by
\beq
\frac{\partial}{\partial \lambda} P(x,y;\lambda)=\int_{y}^{x}
dsP(x,s;\lambda)\frac{\partial}{\partial
\lambda}A(s;\lambda)P(s,y;\lambda).
\eeq

\section{The Non-abelian Stokes Theorem}
To provide the background for using the product integral
formalism of Section II to prove the non-abelian Stokes theorem,
we begin with a statement of the problem as it arises in a
physical context.
Let $M$ be an n-dimensional
manifold
representing the space-time (target space). Let $A$ be a
(connection)
1-form on $M$. When $M$ is a differentiable manifold, we can
choose a
local basis $dx^{\mu}$, $\mu=1,...,n$, and express $A$ in terms
of its
components:
$$
A(x) = A_{\mu}(x) \; dx^{\mu}.
$$
We take $A$ to have values in the Lie-algebra, or a
representation thereof,
of a Lie group. Then, with $T_k$, $k=1,..,m$, representing the
generators
of the Lie group, the components of $A$ can be
written as
$$
A_{\mu}(x) = A_{\mu}^k (x)\; T_k.
$$
With these preliminaries, we can express
the path ordered phase factor of the non-abelian gauge theories
~\cite{mandl,yang,wilson,polyakov} in the
form
$$
W_{ab}(C) = {\cal P} e^{\int_{a}^{b} A}.
$$
where ${\cal P}$ indicates path ordering, and $C$ is a path in
$M$. When the path $C$ is closed, the corresponding holonomy
operator becomes:
\beq
W(C) = {\cal P} e^{\oint A}.
\eeq
The path $C$ in $M$ can be described in terms of an intrinsic
parameter $\s$, so that for points ${x^{\mu}}$ of $M$ which lie
on the
path $C$, we have $x^{\mu} = x^{\mu}(\s)$. One can then write
$$
A_{\mu}(x(\s))dx^{\mu} = A(\s)d\s ,
$$
where
$$
A(\s) \equiv A^{\mu}(x(\s))\frac{dx^{\mu}(\s)}{d\s}.
$$
It is the quantity $A(\s)$, and the variations thereof, which we
will
identify with the matrix
valued functions of the product integral formalism.

Let us next consider the loop operator. For simplicity, we
assume that
$M$ has trivial first homology group with integer coefficients,
i.e.,
$H_1(M,{\bf Z})=0$. This
insures that the
loop may be taken to be the boundary of a two dimensional surface
$\Sigma$
in $M$. More explicitly, we take the 2-surface to be an
orientable
submanifold of $M$. It will be convenient to describe the
properties of
the 2-surface in terms of its intrinsic parameters $\s$ and $\t$
or $\s^a$,
$a=0,1$. So, for the points of the manifold $M$, which lie on
$\Sigma$, we
have $x = x(\s, \t)$. The components of the 1-form $A$ on
$\Sigma$ can be
obtained by means of the vielbeins (by the standard pull-back
construction):
$$
\it{v^{\mu}_a} = \partial_a \; x^{\mu}(\s).
$$
Thus, we get
$$
A_a = \it{v^{\mu}_a} \; A_{\mu}.
$$
The curvature 2-form $F$ of the connection $A$ is given by
$$
F = dA + A \wedge A ={1\over 2} F_{\mu \nu} \; dx^{\mu}\wedge
dx^{\nu}.
$$
The components of $F$ on $\Sigma$ can again be obtained by means
of the
vielbeins:
$$
F_{ab} = \it{v^{\mu}_a} \; \it{v^{\nu}_b}\;F_{\mu \nu}.
$$

We want to express the loop operator in terms of the product
integral definition in a specific way. To achieve this,
we begin with the
definition of the path ordered phase factor in terms of a product
integral.
Consider the continuous map
$A:[s_0,s_1]\rightarrow {\bf C}_{n\times n}$
where $[s_0,s_1]$ is a real interval. Then, we define the
non-abelian phase factor given above in terms of a product
integral as
follows:
$$
{\cal P} e^{\int_{s_0}^{s_1} A(s)ds}
\equiv \prod_{s_0}^{s_1} e^{A(s)ds}.
$$
In particular,
anticipating that we will identify the closed path $C$ over which
the Wilson loop is defined with the boundary of a 2-surface, it
is convenient to work from the beginning with the matrix valued
functions $A(\s, \t)$. This means that our
expression for the path ordered phase factor will depend on a
parameter. That
is, let
\beq
A:[\s_0,\s_1]\times[\t_0,\t_1]\rightarrow {\bf C}_{n\times n},
\eeq
where $[\s_0,\s_1]$ and $[\t_0,\t_1]$ are real intervals on the
two surface $\Sigma$ and hence in $M$.
Then, we define the path ordered phase factor
\beq
P(\sigma,\s_0;\t)= \prod_{\s_0}^{\s} e^{
A_1(\sigma';\t)d\sigma'}
\equiv {\cal P} e^{\int_{\s_0}^{\s} A_1(\sigma';\t)d\sigma'}.
\label{dp}
\eeq
In this expression, ${\cal P}$ indicates path ordering with
respect to $\s$, as defined by the product integral, while $\t$ is 
a parameter. To be able to
describe such an operator for a closed path, we similarly define
the path dependent operator
\beq
Q(\sigma;\t,\t_0) = \prod_{\t_0}^{\t} e^{ A_0(\sigma;\t')d\t'}
\equiv {\cal P} e^{\int_{\t_0}^{\t} A_0(\s;\t')d\t'}.
\label{dq}
\eeq
In this case, the path ordering is with respect to $\t$, and $\s$
is a parameter.

To prove the non-abelian version of the Stokes theorem, we
want to make use of product integration techniques to express the
holonomy loop operator as an integral over a two dimensional
surface bounded by the corresponding loop. In terms of the
intrinsic coordinates of such a surface, we can write this
loop operator in the form
\beq
W(C) = {\cal P} e^{\oint A_a d\s^a},
\label{s24}
\eeq
where, as mentioned above,
\beq
\s^a=(\t,\s)\, ; \quad a=(0,1).
\eeq
The expression for the loop operator depends only on the homotopy
class of
paths in $M$ to
which the closed path $C$ belongs. We can, therefore,
parameterize the path $C$ in any convenient manner consistent
with its homotopy class. In particular, we can break up the path
into piece wise continuous segments along which either $\s$ or
$\t$ remains constant.
The composition rule for product integrals given by Eq. (8)
ensures that this break up of the closed loop into a number of
segments
does not depend on the intermediate points on the closed path,
which are used for this purpose. So,
we write the closed loop operator as
\beq
W = W_4\, W_3\, W_2\, W_1,
\label{w}
\eeq
In this expression, $W_k$, $k=1,..,4$, are Wilson lines such that
$\t = const.$ along $W_1$ and $W_3$, and $\s = const.$ along
$W_2$ and $W_4$. We emphasize that the expressions
$\s = const.$ and $\t = const.$ represent arbitrary
curves.

To see the advantage of parameterizing the closed path in this
manner,
consider the  exponent of Eq. (24) :
\beq
A_a d\s^a = A_0 d\t + A_1 d\s .
\eeq
Along each segment, one or the other of the terms on the right
hand side vanishes. For example, along the segment $[\s_0 ,\s]$,
we have $\t^{'} = \t_0 = const.$.
Recalling Eqs. (22) and (23), we get for the segments $W_1$ and
$W_2$, respectively,
\beq
W_1 = \prod_{\s_0}^{\s} e^{ A_1(\sigma'; \t_0)d\sigma'}
\equiv {\cal P} e^{\int_{\s_0}^{\s}  A_1(\sigma'; \t_0)d\sigma'}
=P(\s,\s_0; \t_0),
\label{sw1}
\eeq
and
\beq
W_2 = \prod_{\t_0}^{\t} e^{A_0(\s; \t')d\t'}
\equiv {\cal P} e^{ \int_{\t_0}^{\t} A_0(\s; \t')d\t'}
=Q(\s; \t,\t_0).
\label{sw2}
\eeq

When the 2-surface $\Sigma$ requires more than one coordinate
patch to cover it, the connections in different coordinate
patches must be related to each other in their overlap region by
transition functions~\cite{yang}. Then, the decomposition given
in Eq. (26) must be suitably
augmented to take this complication into account. The product
integral representation of the path ordered phase factor and the
composition
rule
for product integrals given by Eq. (8) will still make it
possible
to describe the corresponding loop operator as a composite
product
integral.
For definiteness, we will confine ourselves to the representation
given by Eq. (26).

It is convenient for later purposes to define two composite
Wilson line operators
$U$ and $T$ according to
\beq
U(\s, \t) = Q(\s; \t,\t_0)\, P(\s,\s_0; \t),
\label{s30}
\eeq
\beq
T(\s; \t) = P(\s,\s_0; \t)\, Q(\s_0; \t,\t_0).
\label{st}
\eeq
Using the first of these, we have
\beq
W_2\, W_1 = U(\s, \t) .
\eeq
Similarly, we have for $W_3$ and $W_4$
\beq
W_3 = P^{-1}(\s,\s_0; \t),
\eeq
and
\beq
W_4 = Q^{-1}(\s_0; \t,\t_0).
\eeq
From the Eq. (31), it follows that
\beq
W_4\, W_3=T^{-1}(\s,\t).
\eeq
Appealing again to Eq. (8) for the composition of product
integrals, it is clear that this expression for the Wilson loop
operator is
independent of the choice of the point ($\s,\t$).
In terms of the quantities $T$ and $U$, the closed loop operator
will take the compact form
\beq
W=T^{-1}(\s;\t) U(\s;\t).
\eeq

As a first step in
the proof of the non-abelian Stokes theorem, we obtain the action
of the $L$-derivative operator on $W$:
\beq
L_\t W= L_\t[T^{-1}(\s,\t)Q(\s;\t,\t_0) P(\s,\s_0;\t_0)].
\eeq
Using the
definition of the $L$-operation given by Eq. (12), noting that
$P(\s,\s_0;\t_0)$ is
independent of $\t$, and carrying out the $L$ operations on the
right hand side (RHS), we get
\begin{eqnarray}
L_\t W&=& L_\t T^{-1}(\s,\t)+T^{-1}(\s,\t)\,[\,L_\t
Q(\s;\t,\t_0)+
\nonumber\\
&&+Q(\s;\t,\t_0)
(L_\t P(\s,\s_0;\t_0))Q^{-1}(\s;\t,\t_0)]T(\s,\t).
\end{eqnarray}
Simplifying this expression by means of Eqs. (12) and (13), we
end up with
\beq
L_\t W= T^{-1}(\s,\t)[A_0(\s,\t)-L_\t T(\s,\t)]T(\s,\t).
\eeq
Next, we prove the analog of Eq. (14), which applies to an
elementary product integral, for the composite loop operator
defined by Eqs. (26) and (36).

\begin{thm}
\label{thm1}

The loop operator given by Eq. (36) can be expressed
in the form
\beq
W=\prod_{\t_0}^{\t} e^{ T^{-1}(\s, \t')[A_0(\s, \t')  - L_\t
T(\s, \t')] T(\s, \t')d\t'}.
\eeq
\end{thm}
To prove this theorem, first we note from the definition of the
$L$ operation that the right hand side (RHS) of this equation can
be
written as
\beq
RHS =
\prod_{\t_0}^{\t} e^{[
T^{-1}(\s;\t')A_0(\s;\t')T(\s;\t')-T^{-1}(\s;\t'){\partial \over
{\partial \t '}}T(\s;\t')] d\t' }.
\eeq
Noting that $-T^{-1}\partial_\t T=L_\t T$, we can use
the similarity theorem
given by Eq. (17) to obtain
\beq
RHS =
T^{-1}(\s;\t)\prod_{\t_0}^{\t} e^{A_0(\s;\t')d\t'} T(\s;\t_0).
\eeq
Moreover, making use of the defining Eq. (23), we get
\beq
RHS =
T^{-1}(\s;\t)\,Q(\s;\t,\t_0)\,P(\s,\s_0;\t_0)\,Q(\s;\t_0,\t_0)
=T^{-1}(\s;\t)\,U(\s;\t).
\eeq
The last line is clearly the expression for $W$ given by Eq.
(36).

Finally, we want to
express
the quantity $W$ in yet another form which we state as:

\begin{thm}
\label{thm2}

The loop operator defined in Eq. (36) can be expressed
as a surface integral of the field strength:
\beq
W=\prod_{\t_0}^{\t} e^{\int_{\s_0}^{\s} T^{-1}(\s';\t') F_{01}
(\s';\t')T(\s';\t')d\s' d\t'}
\eeq
where $F_{01}$ is the 0-1 component of the non-Abelian field
strength.
\end{thm}
To prove this theorem, we note that
\beq
{\partial \over {\partial
\s}}[T^{-1}(\s,\t)A_0(\s,\t)T(\s,\t)]=
T^{-1}(\s,\t)\left[\partial_\s A_0(\s,\t)
+[A_0(\s,\t), A_1(\s,\t)]\right]T(\s,\t).
\eeq
Moreover,
\beq
{\partial \over {\partial \s}}\{T^{-1}(\s,\t)\left(L_\t
T(\s,\t)\right)T(\s,\t)\} =
T^{-1}(\s,\t)\partial_\t A_1(\s,\t)T(\s,\t).
\eeq
It then follows that
\begin{eqnarray}
&{\partial \over {\partial \s}}\{T^{-1}(\s,\t)[A_0(\s,\t)-L_\t
T(\s,\t)]\}T(\s,\t)\}
\nonumber\\
&= T^{-1}(\s,\t)[{\partial \over {\partial
\s}}A_0(\s,\t)-{\partial \over {\partial \t}}A_1(\s,\t)
+[A_0(\s,\t),
A_1(\s,\t)]\}T(\s,\t)
\nonumber\\
&= T^{-1}(\s,\t)F_{01}(\s,\t)T(\s,\t).
\end{eqnarray}
The last step follows from the definition of the field strength
in terms of the connection given above
\beq
F_{0\,1} ={\partial \over {\partial
\s}}A_0(\s,\t)-{\partial \over {\partial \t}}A_1(\s,\t)
+[A_0(\s,\t),
A_1(\s,\t)].
\eeq
Integrating Eq. (47) with respect to $\s$, we get
\begin{eqnarray}
T^{-1}(\s,\t)[A_0(\s,\t)-L_\t T (\s,\t)]\}T(\s,\t)
\nonumber\\
= \int_{\s_0}^{\s} T^{-1}(\s';\t') F_{0\,1}
(\s';\t') T(\s';\t')d\s' d\t'.
\end{eqnarray}
We thus arrive at the surface integral representation of the
loop operator~\cite{r20}:
\beq
W=\prod_{\t_0}^{\t} e^{\int_{\s_0}^{\s} T^{-1}(\s';\t') F_{0\,1}
(\s';\t') T(\s';\t')d\s' d\t'}.
\eeq
We note that in this expression the ordering of the operators is
defined with respect to $\t$ whereas $\s$ is a parameter.
Recalling the antisymmetry of the components of the field
strength, we can rewrite this expression in terms of path ordered
exponentials familiar from the physics literature:
\beq
W={\cal P_\t} e^{ {1\over 2}\int_\Sigma
d\s^{ab}T^{-1}(\s;\t)F_{ab}
(\s;\t) T(\s;\t)},
\eeq
where $d\s^{ab}$
is the area element of the 2-surface.
Despite appearances, it must be remembered that $\s$ and $\t$
play very different roles in this expression.

\section{A Second Proof}
To illustrate the power and the flexibility of the \p\ formalism,
we give here a variant of the previous proof for the
non-Abelian Stokes
theorem. This time the proof makes essential use of the
non-trivial relation (19) for product integrals.
We start with the form of $W$ given in Eq. (36) and take
 its derivatives with respect to $\t$:
\begin{eqnarray}
{\partial W \over {\partial \t}}&=&\partial_\t
Q^{-1}(\sigma_0;\t,\t_0)
P^{-1}(\s,\s_0;\t) Q(\s;\t,\t_0) P(\s,\s_0;\t_0)+
\nonumber\\
&&+Q^{-1}(\sigma_0;\t,\t_0)
\partial_\t P^{-1}(\s,\s_0;\t) Q(\s;\t,\t_0) P(\s,\s_0;\t_0)+
\nonumber\\
&& +
Q^{-1}(\sigma_0;\t,\t_0)P^{-1}(\s,\s_0;\t) \partial_\t
Q(\s;\t,\t_0)
P(\s,\s_0;\t_0).
\end{eqnarray}
Here, we have made use of the fact that $P(\s,\s_0;\t_0)$ is
independent
of $\t$. Applying
Eq. (12) to $W$ and using Eq. (31), we get
\begin{eqnarray}
L_\t W={\partial W \over {\partial \t}}
W^{-1}=&T^{-1}(\s;\t)\,[A_0(\s;\t)-
P(\s,\s_0;\t)A_0(\s_0;\t)P^{-1}(\s,\s_0;\t)-
\nonumber\\
&-\partial_\t
P(\s,\s_0;\t)P^{-1}(\s,\s_0;\t)]\,T(\s;\t).
\label{n1}
\end{eqnarray}
Now we can use Eq. (19) to evaluate the derivative of
the \p\
with respect to the parameter $\t$:
\beq
\partial_\t P(\s,\s_0;\t)=\int_{\s_0}^\s
d\s'P(\s,\s';\t)\partial_\t A_1
(\s';\t)P(\s',\s_0;\t).
\eeq
Then, after some simple manipulations using the defining
equations
for the various terms in Eq. (53), we get:
\beq
T^{-1}(\s;\t)\partial_\t
P(\t)P^{-1}(\t)T(\s;\t)=\int_{\s_0}^\s d\s'
T^{-1}(\s';\t)\partial_\t A_1
(\s';\t)T(\s';\t).
\label{n2}
\eeq
Using Eq. (9) and the fact that $P(\s_0,\s_0;\t)=1$,
we can write
the rest of
Eq. (53) as an integral too:
\begin{eqnarray}
&T^{-1}(\s;\t)[A_0(\s;\t)-
P(\s,\s_0;\t)A_0(\s_0;\t)P^{-1}(\s,\s_0;\t)]T(\s;\t)=
\nonumber\\
&=Q^{-1}(\s_0;\t,\t_0)[P^{-1}(\s,\s_0;\t)A_0(\s;\t)P(\s,\s_0;\t)
-A_0(\s_0)]Q(\s_0;\t,\t_0)=
\nonumber\\
&=\int_{\s_0}^\s d\s'\, P^{-1}(\s',\s_0;\t)(\partial_\t
A_0(\s',\t
)+[A_0(\s',\t),A1(\s',\t)])P(\s',\s_0;\t).
\end{eqnarray}
Combining Eqs. (53), (55), and (56), we obtain:
\beq
L_\t W={\partial W \over {\partial \t}}W^{-1}=\int_{\s_0}^\s d\s'
T^{-1}(\s',\t)F_{01}(\s',\t)T(\s',\t).
\eeq
Finally, recalling Eq. (14), we are immediately led to Eq. (50)
which was
obtained  by the previous method of proof.

There are two reasons for the relative simplicity of this proof
over the one which was given in the previous section. One is due
to the use of differentiation with respect to a parameter
according the Eq. (19). The other is due to the use of Eq. (14)
for the composite operator $W$. In the first proof,
the use of this theorem for $W$ was not assumed. Its
justification for using it in the second proof lies in the
composition law for product integrals given by Eq. (8).

\section{Gauge Covariance of the result}
As a consistency check, we must show that the surface
representation given by Eq. (50) is gauge covariant. To this end,
it will be recalled that 
under a gauge transformation, the components of the connection,
i.e. the gauge potentials, transform according
to~\cite{peskin}
\beq
A_\mu (x) \longrightarrow g(x)A_\mu (x)g^{-1}(x)-g(x)\partial_\mu
g(x)^{-1}.
\label{gt}
\eeq
The components of the
field strength
(curvature) transform covariantly:
\beq
F_{\mu\nu}(x) \longrightarrow g(x)F_{\mu\nu}(x) g^{-1}(x).
\label{gt2}
\eeq
From these, it follows that~\cite{peskin}
\beq
{\cal P}e^{\int_a^b A_\mu (x) dx^\mu}\longrightarrow g(b)\left(
{\cal
P}e^{\int_a^b A_\mu (x) dx^\mu}\right)g^{-1}(a),
\label{gt1}
\eeq
Equivalently, from its difinition (22) in terms of product
integrals, it is easy to show that the gauge transform of the
quantity $P(\s,\s_0;\t)$ has the form
\beq
g(\sigma;\t)g^{-1}(\sigma_0;\t)\prod_{\s_0}^{\s}
e^{g(\sigma_0;\t)A_1(\sigma';\t)g^{-1}(\sigma_0;\t)}.
\eeq

To show the gauge covariance of the surface representation,
 we need to know how the operator
$T(\s,\t)$ transforms under gauge transformations. To this end,
we note that the Wilson line $Q(\sigma;\t,\t_0)$ given by Eq.
(23) transforms as
\beq
\label{factorq}
Q(\sigma;\t,\t_0) = \prod_{\t_0}^{\t} e^{
A_0(\sigma;\t')d\t'}\longrightarrow
g(\sigma;\t)Q(\sigma;\t,\t_0)g^{-1}(\sigma;\t_0).
\eeq
Then, the gauge transform of the composite operator $T(\s,\t)$
given by
Eq. (31) follows immediately:
\beq
T(\s; \t) = P(\s,\s_0; \t)\, Q(\s_0; \t,\t_0)\longrightarrow
g(\sigma;\t)T(\s; \t) g^{-1}(\sigma_0;\t_0).
\eeq

From the above results, it is straight forward to show that the
surface integral representation of Wilson loop transforms as
\beq
W \longrightarrow
\prod_{\t_0}^{\t} e^{g(\sigma_0;\t_0)\left(\int_{\s_0}^{\s}
T^{-1}(\s';\t') F_{01}(\s';\t')T(\s';\t') dt'\right)
g^{-1}(\sigma_0;\t_0)}.
\eeq
It follows from the composition rule (8) that the constant
factors in the
exponent factorize, so that under gauge transformations the
surface
representation of transforms covariantly, i.e.,
\beq
W \longrightarrow
g(\sigma_0;\t_0)\prod_{\t_0}^{\t}
e^{\int_{\s_0}^{\s} T^{-1}(\s';\t')
F_{01}
(\s';\t')T(\s';\t') dt'} g^{-1}(\sigma_0;\t_0).
\eeq
We view this result as a nontrivial confirmation of our proofs.

\section{Concluding Remarks}
We have provided two proofs of the non-abelian Stokes theorem
using the product integral method. An immediate question which
comes to mind is whether there is a supersymmetric generalization
of this theorem. Given the important developments in
supersymmetric gauge theories in recent years, this question is
not merely of academic interest. To explore this possibility
using the product integral method, it is necessary to generalize
this method to encompass Grassmann valued operators. It turns out
that such a generalization is indeed possible~\cite{km}. Further
developments of this subject will be reported in a forthcoming
work.

\vspace{0.3in}

This work was supported in part by the Department of Energy under
the contract number DOE-FGO2-84ER40153. The hospitality of Aspen
Center for Physics in the Summer of 1998 is also gratefully
acknowledged. We are also grateful to Dr. M. Awada for valuable
input at the initial stages of this work.

\end{document}